\documentclass{elsart}
\usepackage{graphicx}
\usepackage{amssymb}

\begin{document}
\begin{frontmatter}

\hfill DESY 05-098\\
\hfill 11 July 2005\\

\title{Design study of an optical cavity for a future photon-collider
at ILC}

\author[Desy,MBI]{G. Klemz\corauthref{cor}},
\ead{klemz@mbi-berlin.de}
\author[Desy]{K. M\"onig},
\author[MBI]{I. Will},

\address[Desy]{Deutsches Elektronen-Synchrotron, DESY-Zeuthen,
\\ Platanenallee 6, 15738 Zeuthen, Germany}
\address[MBI]{Max-Born-Institute for Nonlinear Optics and Short Pulse
Spectroscopy,
\\ Max-Born Str. 2A, 12489 Berlin, Germany}
\corauth[cor]{Corresponding author. Fax: +49.30.6392-1309}

\begin{abstract}
Hard photons well above 100\,GeV have to be generated in a future
photon-collider which essentially will be based on the
infrastructure of the planned {\em International Linear Collider}
(ILC). The energy of near-infrared laser photons will be boosted
by Compton backscattering against a high energy relativistic
electron beam. For high effectiveness, a very powerful lasersystem
is required that exceeds todays state-of-the-art capabilities. In
this paper a design of an auxiliary passive cavity is discussed
that resonantly enhances the peak-power of the laser. The
properties and prospects of such a cavity are addressed on the
basis of the specifications for the European {\em TeV Energy
Superconducting Linear Accelerator} (TESLA) proposal. Those of the
ILC are expected to be similar.
\end{abstract}

\end{frontmatter}

\section{Introduction}
There is a worldwide consensus that the next particle physics
project is a linear accelerator for positron-electron ($e^+e^-$)
collisions in the energy range of 500\,GeV to
$\approx$\,1\,TeV\cite{sci_case:2003}. In Summer of 2004 the {\em
International Committee for Future Accelerators (ICFA)} recommended
to the {\em International Linear Collider Steering Committee}
(ILCSC) an accelerator based on superconducting radio frequency
resonators for setting up the {\em International Linear Collider}
(ILC) \cite{ICFA:2004}. Such a technology has been developed by the
TESLA ({\em TeV Energy Superconducting Linear Accelerator})
collaboration. Besides collisions between $e^+e^-$, a second
interaction point (IP) for collisions between photons as well as
collisions of electrons on photons is foreseen. This second IP is
commonly referred to as the Gamma-Gamma ($\gamma\gamma$) Collider
arm of a linear collider. In this paper we will continue to use the
parameters of the superconducting TESLA machine \cite{tdr_machine},
as the parameters of the ILC are expected to be similar.

\section{Creation and collision of high energy photons}
\begin{figure}[t]
\includegraphics[width=0.9\textwidth]{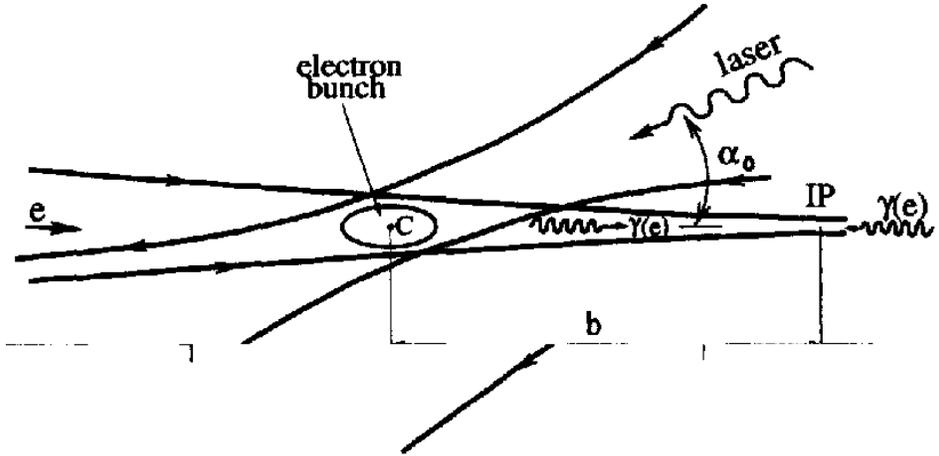}
\caption{Beam geometry around the Compton conversion point $C$ for
creation of highly polarized, high energy photons ($\gamma$)
(taken from \cite{Telnov:2004}). $e$ denotes the 250\,GeV electron
beam. The circular polarized laser has a wavelength of
$\approx$\,1\,$\mu$m. The photons are backscattered into the
flight path of the electrons.}
\label{fig:compton_geometry}
\end{figure}

In a photon collider two opposing pulsed electron beams of
$E_0$\,=\,250\,GeV energy travel towards the  interaction point
(IP). According to the architecture of the TESLA design each of
them collides a few mm before the IP with a tightly focused laser
beam. Its diameter is of the order of a few 10\,$\mu$m, which will
be much larger than the elliptical cross section of the electron
bunch size. The latter exhibits 88\,nm and 4.3\,nm in its two
perpendicular half axes for TESLA \cite{Telnov:2004}. The laser
photons ($\approx1\,$eV) are backscattered as outlined in
Fig.~\ref{fig:compton_geometry}.
Compton scattering raises their energy close to that of the
initial electrons. The produced $\gamma$ propagate in the
direction of flight path of the electrons. There is a small
additional angular spread of the order of
$1/\gamma$\,$\approx$\,2\,$\mu$rad, where
$\gamma=E_0/(m_e\,c_0^2)$ ($m_e$: electron mass, $c_0$: speed of
light in vacuum). Compton backscattering is the most promising
process for efficient creation of highly polarized, $\gamma$ beams
in the range of several hundred GeV with low background
\cite{Telnov:2004,Ginzburg:1981ik,Ginzburg:1983vm,Ginzburg:1984yr}.
At the IP they collide with a similar opposite $\gamma$ or
electron beam. The photon spot size at the IP will then be almost
equal to that of the electrons at the IP.  Therefore the total
luminosity of $\gamma\gamma$, $\gamma e^-$ collisions will be
similar to that of the basic $e^-e^-$ beams.

>From kinematics the maximum photon energy $\hbar\,\omega_m$ is
given by
\begin{equation}
\hbar\,\omega_m=\frac{x}{x+1}E_0\ ,
\qquad
x=\frac{4E_0\, \hbar\omega_0}{m_e^2c_0^4}\cos^2{\frac{\alpha}{2}}
\end{equation}
$E_0$, $\hbar\omega_0$ and $\lambda$ denote the electron beam
energy, photon energy and wavelength of the laser. $\alpha$
represents the crossing angle between laser and electron beam. It
is desirable to keep $x$ below 4.8, since for larger $x$ the high
energy photons are lost by $e^+e^-$ pair production due to their
interaction with the laser beam \cite{Telnov:1995hc}. For the
TESLA beam parameters a wavelength near $\lambda$\,=\,$1\,\mu$m
coinciding with powerful solid state lasers appears to be
promising \cite{Telnov:2004}. Then $x$\,$\approx$\,$4.75$ and
$\hbar\omega_m$\,=\,207\,GeV arise for a 250\,GeV beam. The
resulting Compton spectrum is strongly sensitive to the product of
the mean electron helicity $\lambda_e$
($\mid\!\lambda_{e}\!\mid\le 1/2$) and that of the laser photons
$P_c$\, ($\mid\! P_c\!\mid\le 1$). A value $2\lambda_{e}P_c$ close
to $-1$, i.e. a circular polarized laser beam with opposite
helicity as that of the electrons should be used to maximize
luminosity at high photon energies \cite{Telnov:2004}.

The time structure of the laser system must match the bunch
structure of the accelerator with 2820 bunches per train, each of
2.4\,ps FWHM ($\sigma$\,=\,1\,ps) duration for TESLA. The bunches
are 337\,ns apart and have a 5\,Hz repetition rate
\cite{tdr_machine,Telnov:2004}. Pulses of several Joule are
required to scatter the majority of electrons. The precise energy
of each pulse depends strongly on the degree of focusing the
optical onto the electron beam, as well as on the crossing angle
$\alpha$ in respect to the electrons. A Nd:YLF-based laser
architecture generating this special time structure and providing
a flat and stable train of ultraviolet (UV) synchronized ps pulses
over 0.8\,ms long time periods is already in use for some time at
the TESLA Test Facility (TTF)\cite{SWSLS_NIM:2000,WS_NIM:2004}. At
a wavelength of 1047\,nm it delivers $200\,\mu$J per pulse to the
nonlinear crystals for conversion into the UV. At present, a
maximum single pulse energy in excess of $300\,\mu$J for bursts
containing 800 pulses at 1\,MHz repetition rate, corresponding to
30\,MW peak-power is generated in the infrared. For bursts of 2400
pulses this reduces to $140\,\mu$J per pulse at 3\,MHz repetition
rate within the pulse train \cite{WS_NIM:2004}. This system
produces an average power between one and two watts. On the other
hand, current solid-state laser technology is clearly in the
position of generating short pulses at the  5\,TW peak-power
level. In the foreseeable future however, the required high
average power for a photon collider of several ten kilowatt and
diffraction limited pulses can not be produced directly with the
output from a laser \cite{Meyerhofer:1995,Clayton:hf}.

According to the TESLA machine parameters \cite{Telnov:2004} each
bunch contains $2\,\times\,10^{10}$ electrons, whereas a 5\,J
Laser pulse at $\lambda=1064$\,nm consists of
$2.5\,\times\,10^{19}$ photons. Provided that all electrons within
the bunch undergo scattering, only one in $10^9$ photons is lost
during a single laser-electron collision. Thus, it has been
proposed to reuse the remaining laser pulse by storing it in a
passive resonant optical cavity
\cite{Telnov:2000a,Telnov:2000b,Will:2001ha}. This results in a
significant reduction of the required single pulse energy to be
delivered by the laser. The time structure of the electron bunches
as planned for TESLA is particularly well suited for application
of such a cavity. Its round-trip time has to be adapted to the
bunch-spacing of 337\,ns resulting in a circumference $U$ of about
100\,m. As will be shown, this is sufficient for wrapping the
optical cavity around the particle detector.

\section{Existing applications of passive optical cavities}
Passive build-up cavities for generating a region of enhanced
intensity are routinely used in combination with both continuous
wave (cw) and mode-locked sources in a number of laser-related
experiments such as frequency doubling
\cite{TR03,KAIMO03,FZ01,ABD66} (cw-laser),
\cite{McCFL:2001,PTF90,MF:89} (mode-locked laser), high-resolution
spectroscopy \cite{SDEKRvO97,Damm96,LoHall90,GAJF90,BB89}, high
sensitive detection of absorption
\cite{IA99,EBPM98,RKSS97,NKSLO94} or cavity ring-down absorption
spectroscopy \cite{GR:2002}. For the latter, linear cavities for
micro-pulse energy enhancement in the mid-infrared region at
5.3\,$\mu$m have been investigated at the Stanford free-electron
laser \cite{CHMS99,SHS97}.

A recent application is the amplification of ultrashort light
pulses through phase coherent superposition of successive pulses
from a mode-locked pulse train in a high-finesse optical cavity
and subsequent cavity dumping \cite{VRH:2003,PEXJY:2003,JY:2002}.
An amplification factor in the order of 10 has been obtained.

For optical interferometric detection of gravitational waves a
2000-fold power recycling cavity for the GEO600-project, one of
several current first-generation large-scale interferometers, is
specified to develop a cw light power of approximately 10\,kW within
the interferometer \cite{Lueck:97}. Recently, a power enhancement in
the range between 1000 and 1200 has already been obtained with
approximately 5\,W input from the mode cleaner cavities
\cite{K.Strain:2006}.

Laser increasingly enter also the field of high energy physics for
monitoring the transverse size of electron beams (``laser wire'')%
\cite{Sakai:2002vv,Sakai:2002ud,Sakai:2002udb,Sakai:2001pb,%
Sakai:2000ck,Shintake:1992qt}, as well as for precision
measurement of electron beam polarization
\cite{Beckmann:2000cg,Passchier:1998ff,SLD}. Both are based on
Compton scattering of low energy photons around 1\,eV. The "laser
wire" technique relies on an optical cavity for boosting the laser
power.

Most of the current cavities for the purpose of frequency doubling
have a ring configuration, providing optical isolation from the
laser cavity. This advantage has been pointed out in
\cite{KS:1992}. The ring configuration is also the preferred
geometry for an optical cavity for the photon collider.

Some non-resonant storage rings have been demonstrated in which
the energy of a laser pulse that was lost through scattering,
diffraction at the mirrors and upon impact with the electron beam
is partially replaced by an amplifier \cite{MAS02,MAS04}. In these
cases, the round-trip time of the optical pulse is less than the
intervals of the laser pulses. These storage rings operate well
below the power density required for the $\gamma\gamma$-collider.

\section{Energy storage properties of the passive optical cavity}
A short pulse is a wave packet with a wide frequency spectrum. One
therefore has to pursue the transfer function approach,
whereas the transient response of the circulating field strength
$\mathcal E_{circ}(t)$ within the cavity to the incoming TESLA
burst-mode time structure $\mathcal E_{in}(t)$ is mathematically
described by a convolution of the incoming laser pulses with the
impulse response function $H(t)$ of the cavity \cite{Systems:1968}:
\begin{equation}
\label{general_response} \mathcal E_{circ}(t)\,=\,\mathcal
E_{in}(t)\,\star\,H(t)
= \int_{-\infty}^{+\infty}
\mathcal E_{in}(\tau)\,H(t-\tau)\,d\tau
\qquad\mbox{.}
\end{equation}
Then $\left|\mathcal E_{circ}(t)\right|^2$ is proportional to the
circulating intra-cavity power $\mathcal P_{circ}(t)$ and the
incident power $\mathcal P_{in}(t)$. $A(t)$\,:=\,$\mathcal
P_{circ}(t) / \mathcal P_{in}(t)$ defines the enhancement factor of
the cavity for the incoming power $\mathcal P_{in}(t)$. For sake of
simplicity we assume bursts of $q$ rectangular shaped laser pulses,
each of duration $\tau_L$, that are synchronous with the round-trip
time $\tau_{circ}$ of the cavity. Such a train of pulses is
represented by a convolution of the function describing the
characteristics of a single pulse with the sampling function:
\begin{eqnarray}
\label{burst} \mathcal E_{in}(t) &=& \mathcal
E_{0}\,\e^{i\,2\,\pi\,\nu_0\,t}\:rect_{\displaystyle\tau_L}(t)
\star\
\sum_{j=0}^{q-1} \delta(t-j\,\tau_{circ})\\
\nonumber &=&\mathcal E_{0}\,\e^{i\,2\,\pi\,\nu_0\,t}\;
\sum_{j=0}^{q-1}\, rect_{\displaystyle\tau_L}(t-j\,\tau_{circ})\ ,
\quad\mbox{with}
\end{eqnarray}
\[
rect_{\displaystyle\tau_L}(t)=\left\{
\begin{array}{ll}
\label{bunch}
  1 & \ \mbox{if}\ \ |t| < \tau_L/2\\
  0 & \ \mbox{if}\ \ |t| > \tau_L/2
\end{array}
\right. \quad\mbox{.}
\]
As a consequence of the shifting property of the $\delta$-function,
the circulating field of the cavity from
Eq.~(\ref{general_response}) is a sum of properly shifted and scaled
replicas of the injected field $\mathcal E_{in}(t)$. The convolution
of Eq.~(\ref{general_response}) can however more conveniently be
solved by taking the inverse Fourier transform of the intra-cavity
spectrum $\widetilde{\mathcal E_{circ}(\nu)}$. Then one benefits
from the convolution property of the Fourier transform that states
\begin{equation}
\label{convolution_Theorem} \widetilde{\mathcal E_{circ}(\nu)}\,=
F\left[\mathcal E_{circ}(t)\right]\,=\, F\left[\mathcal
E_{in}(t)\,\star\,H(t)\right]\,=\, F\left[\mathcal
E_{in}(t)\right]\, F\left[H(t)\right] \quad\mbox{.}
\end{equation}
The input spectrum $\widetilde{\mathcal E_{in}}(\nu)$ of the burst
from Eq.~(\ref{burst}) according to the Fourier transform
\begin{equation}
\label{FT_forward} \widetilde{\mathcal E_{in}(\nu)}\,=\,
F\left[\mathcal E_{in}(t)\right]\,= \int_{-\infty}^{+\infty}\mathcal
E_{in}(t)\,\e^{-i\,2\,\pi\,\nu\,t}\,dt
\end{equation}
results in
\begin{eqnarray}
\label{input_spectrum} \widetilde{\mathcal E_{in}(\nu)}&=&
 \mathcal E_{0}\,
 \int_{-\infty}^{+\infty}
 \e^{i\,2\,\pi\,(\nu_0-\nu)\,t}\,
 \sum_{j=0}^{q-1}\,
 rect_{\displaystyle\tau_L}\left(t-j\,\tau_{circ}\right)\,dt
\\\nonumber
&=&
 \mathcal E_{0}\,
 \frac{\sin\left[\pi\,(\nu_0\!-\!\nu)\tau_L\right]}
 {\pi\,(\nu_0\!-\!\nu)}\,
 \exp\left[
 i\,(q\!-\!1)\,\pi\,(\nu_0\!-\!\nu)\tau_{circ}
 \right]\,\\\nonumber
&&\hspace{2.7ex}
 \frac{
 \sin\left[q\,\pi\,(\nu_0\!-\!\nu)\,\tau_{circ}\right]
 }{
 \sin\left[\pi\,(\nu_0\!-\!\nu)\,\tau_{circ}\right]
 }
\quad\mbox{.}
\end{eqnarray}

The transfer function $\widetilde {H(\nu)}$ for the power build-up
ring cavity is almost identical to that of a Fabry-Perot
interferometer. Extending the derivation described in
\cite{Cesini:1977} to a ring cavity with loss the transfer function
can be expressed as
\begin{equation}
\label{Transfer_function} \widetilde{H_{}(\nu)}\,=\,
F\left[H(t)\right]\,=\, \frac{
\sqrt{1-R_c}\,\exp\left(-i\pi\nu\tau_{circ}\right)}
{1-\sqrt{R_c\,V}\,\exp\left(-i\,2\pi\nu\tau_{circ}\right)}
\quad\mbox{.}
\end{equation}
In the above equation $\widetilde{H_{}(\nu)}$ repesents the
amplitude of a monochromatic wave within the cavity at optical
frequency $\nu$ that is due to a monochromatic input field at same
frequency with unit amplitude. $R_c$ represents the intensity
reflectivity of the coupling mirror and $V$ the power loss factor
for one round-trip. The latter subsumes all loss mechanisms:
intra-cavity absorption, diffraction and scattering as well as the
finite reflectivity of all $N$ cavity mirrors with exception of the
coupling mirror. $R_c$ and $V$ are related to the corresponding
quantities for the field amplitudes (denoted with small letters) by
\begin{equation}
\left.
\begin{array}{l}
r_j = \sqrt{R_j}\quad\mbox{,}\qquad j=1\ldots N\\[1ex]
r_{eff}\cdot v = \sqrt{R_c}\, \underbrace{
\sqrt{R_2}\,\cdot\,\ldots\,\cdot\,\sqrt{R_N}\,\,\cdot\,v}_{
 \displaystyle=:\sqrt{V}}
 =\sqrt{R_c\,V}\\
\end{array}
\quad\right\} \qquad.
\end{equation}
As stated in \cite{Cesini:1977} the square modulus of the transfer
function Eq.~(\ref{Transfer_function}) gives the power response to a
unit power monochromatic input field. This turns out to be the Airy
shape function with Finesse $\mathcal F$ as a parameter
\cite{BoWo:1999}:
\begin{equation}
\label{Airy} \left| \widetilde{H_{}(\nu)} \right|^2=
\frac{\displaystyle 1-R_c}{\displaystyle
\left(1-\sqrt{R_c\,V}\right)^2}\ \frac{1}{1+
\left[\frac{\displaystyle 2\,\mathcal F}{\displaystyle \pi}
\sin\left( \pi\,\nu\,\tau_{circ} \right)\right]^2 } \quad,
\quad\mbox{and}
\end{equation}
\begin{equation}
\mathcal F = \frac{\displaystyle
\pi\,\sqrt[4]{R_c\,V}}{\displaystyle
 1-\sqrt{R_c\,V}} \quad.
\end{equation}
The cavity field is maximized at the resonances of the cavity which
occur if $\nu\,\tau_{circ}$ equals a positive integer number $n$. In
terms of the round-trip path length $U$\,=\,$c_0\tau_{circ}$ and
optical wavelength $\lambda$\,=\,$c_0/\nu$ with $c_0$ representing
the speed of light in vacuum, the resonance condition is simply
\begin{equation}
\label{resonance} U = n\,\lambda \quad\mbox{with a positive integer
number $n$\quad.}
\end{equation}

The transfer function describes the evolution of the electric
field at an arbitrarily chosen point within the cavity. in
Eq.~(\ref{Transfer_function}) it is located a propagation distance
of $U/2$ behind the coupling mirror. This can be seen by the
argument $\nu\, \tau_{circ}$ of the complex exponential in the
numerator instead of $2\,\nu\,\tau_{circ}$ for the complete
revolution which appears in the denominator.

\begin{figure}[h]
\includegraphics[width=0.9\textwidth]{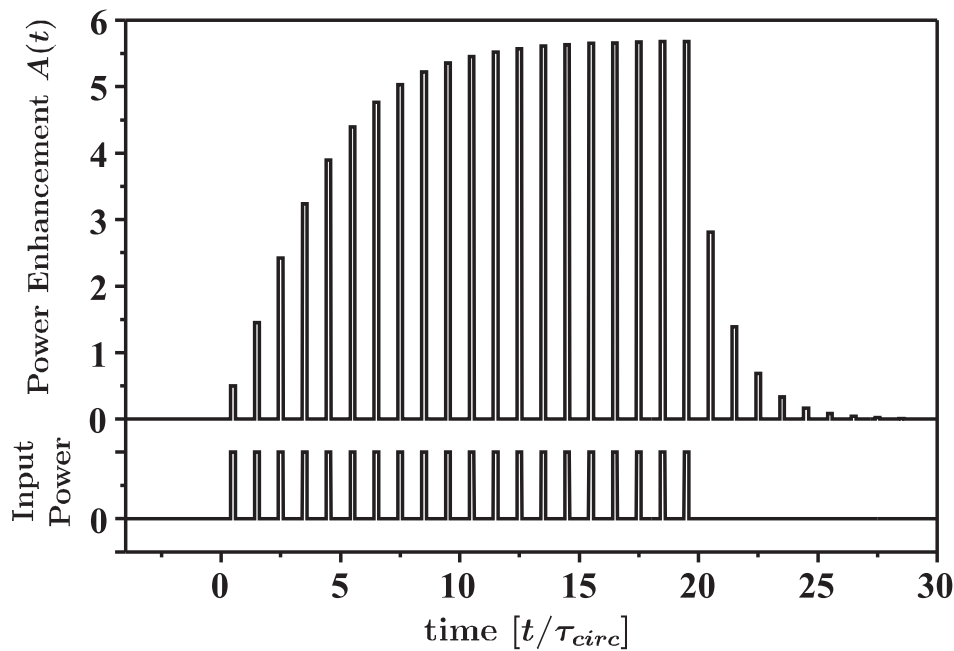}
\caption{Illustration for the evolution of pulse power within the
cavity at a distance $U/2$ away from the coupling mirror ({\sl upper
part}) and incoming burst of 20 lasers pulses at the same position
without presence of the cavity ({\sl lower part}). The duration of a
single pulse is assumed to be $\tau_L$\,=\,$\tau_{circ}/5$. In this
example 11 pulses are needed for the arbitrary chosen parameters
$R_1$=50\,\%, $V$=0.99 in order to reach a power enhancement of at
least 95\,\% of the stationary value $A_{max}\approx 5.7$.
}\label{fig:enhancement_evolution}
\end{figure}

The above functions Eq.~(\ref{input_spectrum},
\ref{Transfer_function}) are combined to the intra-cavity spectrum
$\widetilde{\mathcal E_{circ}(\nu)}$\,=\, $\widetilde{\mathcal
E_{in}(\nu)}\, \widetilde{H_{}(\nu)}$. By subsequent application of
the inverse Fourier transform
\begin{equation}
\label{FT_back} \mathcal E_{circ}(t)\,=\, F^{-1}\left[
\widetilde{\mathcal E_{circ}(\nu)}\right]\,=\,
\int_{-\infty}^{+\infty}\widetilde{\mathcal
E_{circ}(\nu)}\,\e^{i\,2\,\pi\,\nu\,t}\,d\nu
\end{equation}
one finally arrives at the intra-cavity-field
\begin{eqnarray}
\label{E_circ} \mathcal E_{circ}(t) &=&
  \mathcal E_{0}\,
  \sqrt{1-R_c}\,
  \sum_{n=0}^{\infty}\,
  (R_c\,V)^{n/2}\,
  \exp\left[
  2\pi i\,\nu_0 \left\{t-\left(
  n+\frac{1}{2}\right)\tau_{circ}\right\}
\right]\nonumber\\
&&\sum_{j=0}^{q-1}\, rect_{\tau_{L}} \left[t-\left(
n+j+\frac{1}{2}\right) \,\tau_{circ} \right] \quad\mbox{.}
\end{eqnarray}
Since $\left|\sqrt{R_c\,V}\,\exp\left(2\pi i
\nu_0\,t_{circ}\right)\right|<1$, the identity
\begin{equation}
\frac{1}{1-\sqrt{R_c\,V}\,\exp\left( -2\pi i\nu\tau_{circ} \right)}
\,=\, \sum_{n=0}^{\infty} (R_c\,V)^{n/2}\, \exp\left(-2\pi i
\nu\tau_{circ}\right)
\end{equation}
together with properties of the Fourier transform could be used in
derivation of Eq.~(\ref{E_circ}). Since $\tau_{circ}\,\gg\,\tau_L$,
the round-trip time $\tau_{circ}$ introduces a staircase behavior
for the resulting intra-cavity power. Two regimes evolve from
Eq.~(\ref{E_circ}). The power builds up during the first $p$
revolutions of the pulse as long as energy is fed from the burst
mode laser into the cavity ($p\le q$). When all $q$ laser pulses of
the burst had been issued the stored power decays during the
subsequent circulations ($p>q$).
\begin{equation}
\label{generell_Pcirc}
\frac{\displaystyle \mathcal P_{circ}(t)}
     {\displaystyle \mathcal P_{in}(t)} =\left\{
\begin{array}{l}
 (1-R_c)
 \frac{\displaystyle \left(1-\sqrt{R_c\,V}^{\,p}\right)^2+
 4\sqrt{R_c\,V}^{\,p}\,\sin^2(p\pi\nu_0\tau_{circ})}
 {\displaystyle\rule{0ex}{3.6ex} \left(1-\sqrt{R_c\,V}\right)^2+
 4\sqrt{R_c\,V}\,\sin^2(\pi\nu_0\tau_{circ})}\\[1.3em]
\qquad\mbox{if}\quad p\le q\quad
 \mbox{and} \quad
  (2p+1)\,\tau_{circ}\!-\!\tau_L < 2t <
  (2p+1)\,\tau_{circ}\!+\!\tau_L\\[2em]
 (1-R_c)
 \frac{\displaystyle \left(1-\sqrt{R_c\,V}^{\,q}\right)^2+
 4\sqrt{R_c\,V}^{\,q}\,\sin^2(q\pi\nu_0\tau_{circ})}
 {\displaystyle\rule{0ex}{3.6ex} \left(1-\sqrt{R_c\,V}\right)^2+
 4\sqrt{R_c\,V}\,\sin^2(\pi\nu_0\tau_{circ})}
 \left(R_c\,V\right)^{p-q}\\[1.3em]
\qquad\mbox{if}\quad p>q\quad
 \mbox{and} \quad
  (2p+1)\,\tau_{circ}\!-\!\tau_L < 2t <
  (2p+1)\,\tau_{circ}\!+\!\tau_L\\[2em]
  0
 \hspace{1.14ex}\quad \mbox{if $t$ otherwise}\\
\end{array}
\right.
\end{equation}
At resonance the enhancement of the incoming laser pulse power
$\mathcal P_{in}(t)$ simplifies to
\begin{equation}
\label{resonant_Pcirc}
\frac{\displaystyle \mathcal P_{circ}(t)}{\displaystyle
\mathcal P_{in}(t)} =\left\{
\begin{array}{ll}
 (1-R_c) \left[
 \frac{\displaystyle 1-\sqrt{R_c\,V}^{\,p}}
      {\displaystyle 1-\sqrt{R_c\,V}}
 \right]^2
& \quad \mbox{if}\quad p\le q\quad \mbox{and}\\[1.3em]
\multicolumn{2}{r} {    (2p+1)\,\tau_{circ}\!-\!\tau_L < 2t <
     (2p+1)\,\tau_{circ}\!+\!\tau_L
     }\\[2em]
 (1-R_c) \left[
 \frac{\displaystyle 1-\sqrt{R_c\,V}^{\,q}}
      {\displaystyle 1-\sqrt{R_c\,V}}
 \right]^2
 \left(R_c\,V\right)^{p-q}
& \quad \mbox{if}\quad p> q\quad \mbox{and}\\[1.3em]
\multicolumn{2}{r} {    (2p+1)\,\tau_{circ}\!-\!\tau_L < 2t <
     (2p+1)\,\tau_{circ}\!+\!\tau_L
     }\\[2em]
  0 \hspace{4.6em}\mbox{if $t$ otherwise}\\
\end{array}
\right.
\end{equation}
Fig.~\ref{fig:enhancement_evolution} shows the typical transient
behavior of the power within the cavity according to
Eq.~(\ref{resonant_Pcirc}).

For an uninterrupted pulse train ($p\rightarrow\infty$) the
enhancement factor $A_p$ at completion of $p$ round-trips
converges to:
\begin{equation}
\label{Ehf_max} A_{max} = \frac{\displaystyle 1-R_c}{\displaystyle
\left(1-\sqrt{R_c\,V}\right)^2} \qquad.
\end{equation}
The same expression arises from a monochromatic, i.e. continuous
wave. An efficient enhancement will require $R_c$ and $V$ close to
one. Stationary conditions will then be reached to a very good
approximation within each burst of the TESLA time structure.

Besides power reflectivities and losses the power enhancement is
determined by the length detuning $\delta$ of the cavity. Any
arbitrary circumference can be expressed as
$U$\,=\,$n\,\lambda+\delta$ with $\delta < \lambda$. From
Eq.~(\ref{generell_Pcirc}) follows then for sufficient large $p$:
\begin{equation}
\label{Ehf} A(\delta)\approx A_{max} \frac{1}{1+
\left[\frac{\displaystyle 2\,\mathcal F}{\displaystyle \pi}
\sin\left(\pi\,\frac{\displaystyle \delta}{\displaystyle
\lambda}\right)\right]^2 } \quad.
\end{equation}

\begin{figure}[t]
\includegraphics[width=0.9\textwidth]{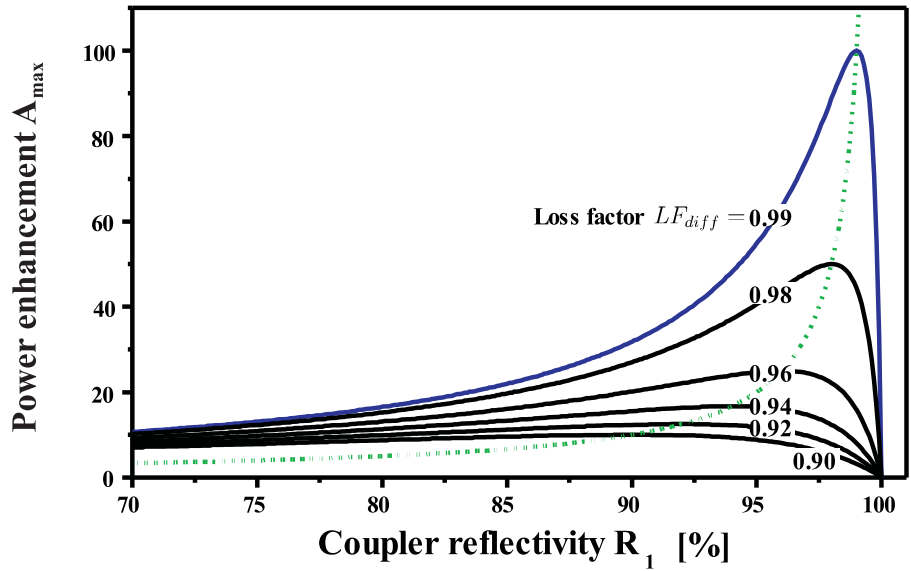}
\caption{The computed enhancement factor $A_{max}$ as a function
of the input mirror reflectivity $R_c$ at different values of the
power loss factor $V$ according eq.\,(\ref{Ehf_max}). Note the
preferable situation for impedance matching (dotted trace).
$R_1$ denotes the reflectivity of the coupling mirror.}
\label{fig:simple_enhancement}
\end{figure}

If the reflectivity of the coupling mirror equals a given loss
factor $V$, the power enhancement takes its maximum possible value
$1/(1-V)$ and all light is absorbed by the resonant cavity. This is
known as impedance matching ($R_c:=V$) in analogy to properties of
electrical cables and waveguides. In
Fig.~\ref{fig:simple_enhancement} the dependence of the enhancement
factor $A$ on the input mirror reflectivity at different values of
$V$ is plotted. Each curve shows a maximum value $1/(1-V)$  for
$R_c$ equal to $V$.

The resonant ($\delta=0$), impedance-matched power enhancement can
also be expressed in terms of the finesse $\mathcal F$ given by
\begin{equation}
\label{eq:enhance-Finesse}
\frac{\displaystyle \mathcal F}{\displaystyle \pi}
=\sqrt{A_{max}\,(A_{max}-1)}
\approx A_{max}\qquad (A_{max} \gg 1)
\end{equation}
Hence, a Finesse in excess of 300 corresponds to an assumed
example of $A_{max}$\,=\,100.

The duration for which a laser pulse will be stored within the
cavity is given by the cavity photon lifetime $\tau_{cav}$. That is
defined by the approximately exponential decay of the power
following a sudden turn-off of the injected laser pulses at some
time $t_0$. Then $\tau_{cav}$ turns out to be represented by
\begin{equation}
\label{cavity_lifetime_2}
\tau_{cav} =
 -\,\tau_{circ}\,
 \frac{\displaystyle 1}{\displaystyle  2\,\ln\left(1-
   \frac{\displaystyle 1}{\displaystyle A_{max}} \right)}
\approx
 \tau_{circ}\,
\frac{\displaystyle A_{max}}{\displaystyle 2}
 \quad (A_{max}\gg 1)
\end{equation}
for the impedance matched cavity. A round-trip time of
$\tau_{circ}$\,=\,337\,ns and $A_{max}$\,=\,100 result in
$\tau_{cav}$\,$\approx$\,49.75$\,\times\,$337\,ns\,%
$\approx$\,16.8\,$\mu$s. Using this result the effective number
$n_{rt}$ of round-trips of an optical pulse within the cavity is
determined by
\begin{equation}
n_{rt}= \frac{\displaystyle \tau_c}{\displaystyle \tau_{circ}}=
-\frac{\displaystyle 1}{\displaystyle
\ln\left(R_c\,V\right)}\approx \frac{\displaystyle
A_{max}}{\displaystyle 2} \approx \frac{\displaystyle \mathcal
F}{\displaystyle 2\,\pi} \qquad.
\end{equation}
This number amounts to $n_{rt}$\,$\approx$\,50 in the above
example.

It can be shown that for an uninterrupted pulse train the power that
is reflected off an impedance matched cavity ($R_c\,=\,V$) declines
as
\begin{equation}
\mathcal P_{reflec}^{(p)}=V^{2\,p-1}\, \,\mathcal P_{in}
\quad\longrightarrow\ 0\quad(p \rightarrow \infty) \qquad.
\end{equation}
No power will thus be reflected from an impedance matched cavity in
steady state. This behavior can be used as an indicator for the
alignment of the cavity in an automatic control system.

\section{Constraints due to the particle detector and design proposal
for an optical cavity}
The Compton-interaction requires operation of the cavity in the
Ultra High Vacuum (UHV) of the accelerator. When operated at the
inevitable high power level optical windows within the cavity
would imply the risk of distortion of the circulating optical
ps-pulse as a result of the non-vanishing B-integral
\cite{Sieg_b:1986,Koe:1992}. For maintaining a sufficient high
$\gamma$ flux density and hence a luminosity of around
$10^{34}$\,cm$^{-2}$\,s$^{-1}$, the laser pulse must be focused at
the Compton conversion point (CP).

The particle detector for the $\gamma\gamma$-option is expected to
be almost identical to the one for $e^+e^-$-physics located at the
leptonic IP. For TESLA, the path length between its end face and
the focus of both optical cavities amounts to about 7.40\,m
\cite{Meyners:2001dz}. The end faces extend over 7.45\,m and are
perpendicular to the beam axis. As a consequence of the required
tight optical focus, a widespread beam diameter of more than half
a meter will emerge from the detector end-faces. Due to shortage
of space any optics for the optical cavity should preferably be
positioned outside the environment of the particle detector.

These demands suggest to employ two symmetrical telescopic ring
cavities, required for Compton conversion at each of the
counter-propagating electron beams, and to interlace them without
mutual interference. Fig.~\ref{fig:embedded-cavities} depicts an
aerial view on two possible spatial embeddings of these cavities.
{\unitlength1mm
\thicklines
\begin{figure}[t]
\includegraphics[width=0.9\textwidth,bb=58 24 1044 745,clip]{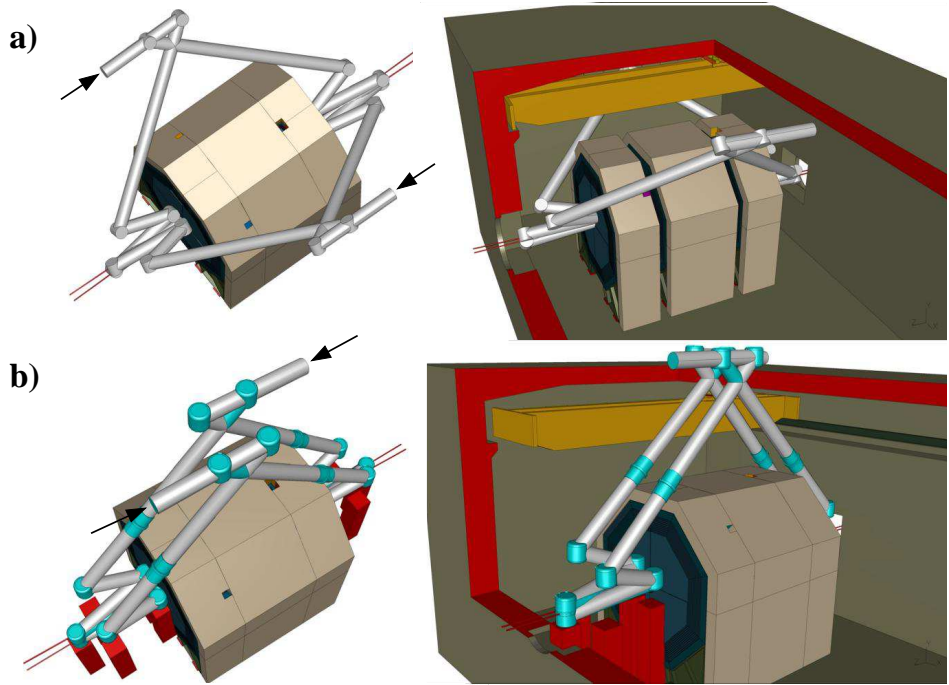}
\caption{Schematic aerial view on two possible configurations for
folding both optical cavities for the TESLA photon collider around
the detector ({\sl left}). Their respective placement in the
experimental hall is also depicted ({\sl right}). The laser beams
are coupled into the cavity at positions marked by the arrows. The
optical beam path is contained within the sketched pipes that
preserve the vacuum. The high power lasers itself will be located in
a separate hall above the detector (not shown). The thin lines
traversing the detector represent the electron beam paths. A slight
mutual vertical tilt between the cavities permits free passage of
the particle beams. {\sl As a ruler:} The detector extends 14.8\,m
along the electron path.} \label{fig:embedded-cavities}
\end{figure}
Their optical paths are enclosed within the associated optical beam
pipes which are needed for maintaining the vacuum. In the upper
sketch each ring resonator stretches across a plane. Both planes
enclose an angle of $90^{\circ}$ degrees. This tipping allows an
un-obstructed passage of both electron beams directions. In
addition, a small tilt angle of approx. $+1.58^{\circ}$ and
$-1.58^{\circ}$, respec., in the vertical direction prevents the
mirrors of the final focusing optics being to close to the leptonic
beams. The two laser beams originate from a separate hall placed
directly above the detector. They are pointing in opposing
directions with the same vertical tilt for coupling into their
respective cavities on top of the detector. For the lower design,
the optical beam paths outside the detector are kept in two adjacent
parallel planes. They cross each other in the interaction region
within the detector.

Fig.~\ref{fig:telescopic_cavity} describes the optical
configuration of an individual cavity.
\begin{figure}[t]
\includegraphics[width=\textwidth]{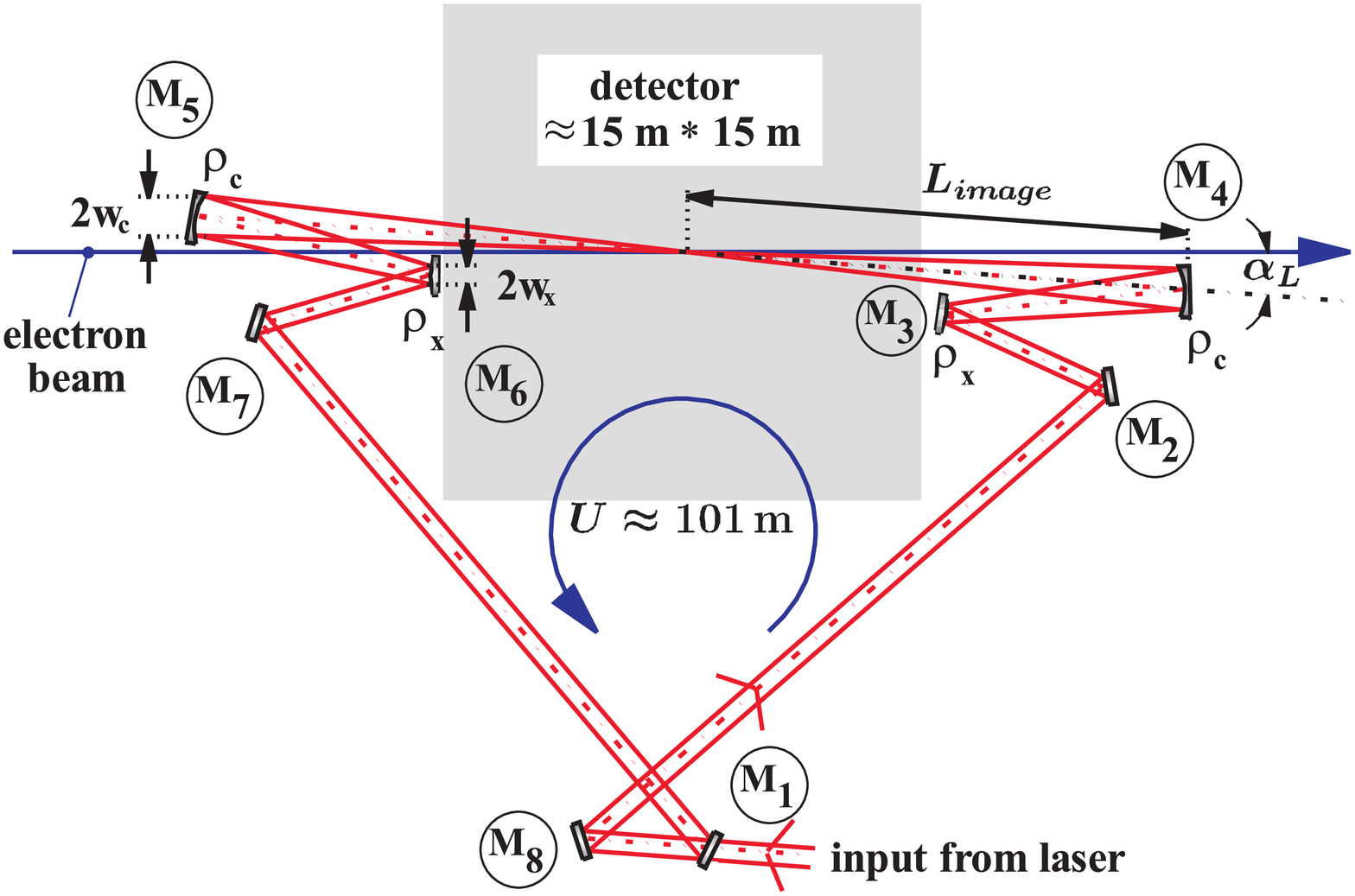}
\caption{Geometry (to scale) of one of the identical plane cavity,
comprising a beam magnification $\mu=w_c/w_x$. CP: Compton
conversion point.} \label{fig:telescopic_cavity}
\end{figure}
A telescopic convex-concave mirror arrangement generates a focus in
the interaction region and a second, identical telescope
re-collimates it again. In addition, they introduce a moderate beam
magnification that reduces the beam size within the nearly
collimated region of the cavity outside the detector. Here, the
laser is coupled to the cavity via M$_1$ or M$_8$.

\section{Telescopic cavity}
Disregarding the final size of the mirrors the focal spot size at
CP is determined by the Gaussian eigenmode within this cavity.
That is calculated by setting up the round-trip matrix
$\left(\left(M_{\mbox{\footnotesize ccw}}\right)\right)$,
\cite{Sieg_c:1986}. We assume total correction of the aberrations
of the telescope mirrors. When starting at a reference plane
containing CP and proceeding in counter clockwise direction, one
obtains for the proposed cavity the following dependence on the
geometrical data:
\begin{eqnarray}
\label{eq:RT-matrix} \left(\left(M_{\mbox{\footnotesize
ccw}}\right)\right)= \left( \begin{array}{cc} G^{\ast} &
\frac{\displaystyle \rho^{\ast}\, g_1^{\ast}}
 {\displaystyle 2}\,(1-G^{\ast})\\
-4\,\frac{\displaystyle g_2^{\ast}}{\displaystyle \rho^{\ast}} &
G^{\ast}\end{array} \right)
\;,\quad\mbox{with these shortcuts}
\end{eqnarray}
\begin{equation}
\label{eq:shortcuts}
\begin{array}{l}
\left. {\begin{array}{l@{\qquad}l}
 G^{\ast} = 2\,g_1^{\ast} g_2^{\ast} -1
  &
 L_1^{\ast}=\frac{\displaystyle L_1}{\displaystyle\rule{0em}{2ex} M}
  + \rho_x \\[1ex]
 g_1^{\ast} =M\,\left(1-\frac{\displaystyle L_1^{\ast}}
  {\displaystyle \rho^{\ast}}
 \right)
  &
 L_2^{\ast}=M\, L_2 + \rho_c \\[2ex]
 g_2^{\ast} =\frac{\displaystyle 1}{\displaystyle\rule{0em}{2ex} M}\,
 \left(1-\frac{\displaystyle L_2^{\ast}}{\displaystyle \rho^{\ast}}
 \right)
  &
 \rho^{\ast}=-\frac{\displaystyle \rho_c\rho_x}
  {\displaystyle\rule{0em}{2ex} 2\,\delta} \\[1ex]
  \\
 M = \left|\frac{\displaystyle f_c}{\displaystyle\rule{0em}{2ex}
  f_x}\right|
  &
 \delta = t - (f_c +f_x), \quad f_c= \rho_c/2, \quad f_x=\rho_x/2\;, \\
 \end{array}}
\ \right\}\\
\end{array}
\,
\end{equation}
This results from multiplication of 2$\times$2-matrices describing
the sequence of reflections on $i$ mirrors with radii of curvature
$\rho_i$ and the free space propagations in between. $L_1$
represents the distance between both concave mirrors $M_4$ and
$M_5$. Between $M_6$ and $M_3$ the ring is closed by $L_2$ via
$M_7$, $M_1$, $M_8$ and $M_2$ of Fig.~\ref{fig:telescopic_cavity} on
the lower path. $t$ designates the length of the telescope (spacing
between $M_3$ and $M_4$ as well as between $M_5$, and $M_6$) and
$\delta_{tel}$ quantifies a detuning from the length $f_c+f_x$ for
afocal alignment of the telescope set up by concave and convex
mirrors of effective focal lengths $f_c$ and $f_x$, respectively.
Due to the inclined angle of incidence of the laser beam on the
curved mirrors their focal lengths are modified, which is expressed
by the effective value. They should have parabolic surfaces for
reducing optical aberrations, predominantly astigmatism
\cite{Kor:1991}. Suitable definitions for a design of the cavity are
the beam magnification $\mu$ as the ratio of beam radii $w_c$ and
$w_x$ on the concave and convex mirror of one telescope, as well as
the final image distance $L_{image}$ between the focus and the
concave mirror of each telescope when a collimated beam enters the
telescope on the side of the convex mirror.
\begin{equation}
\label{eq:telescope}
\begin{array}{l}
\mu:=\frac{\displaystyle w_c}{\displaystyle w_x}
=M-\frac{\displaystyle \delta_{tel}}{\displaystyle\rule{0em}{2ex}
f_x} \qquad\qquad L_{image}=\frac{\displaystyle
\mu\,f_c^2}{\displaystyle\rule{0em}{2ex} \delta_{tel}\,M}
=\frac{\displaystyle f_c\,(t-f_x)}{\displaystyle\rule{0em}{2ex}
t-f_c-f_x}
=\frac{\displaystyle L_1}{\displaystyle\rule{0em}{2ex} 2}\ .\\
\end{array}
\end{equation}
The location of the telescope within the cavity determines the
beam size $w_{CP,\,Gaussian}$ of the Gaussian focus at the
reference plane CP \cite{Sieg_c:1986}:
\begin{equation}
\label{eq:Cavity-Solution-w}
w_{CP,\,Gaussian}^2=\frac{\displaystyle \lambda}{\displaystyle
\pi}\,\rho^{\ast}\, \sqrt{ \frac{\displaystyle
g_1^{\ast}}{\displaystyle g_2^{\ast}}\,
\left(1-g_1^{\ast}\,g_2^{\ast}\right) } \quad.
\end{equation}
For a stable --- i.e. on each revolution reproducing Gaussian beam
--- an additional confinement-condition has to be fulfilled:
\begin{equation}
\label{eq:confinement}
-1 \le G^{\ast} \le 1
\qquad.
\end{equation}
The diameter of the final focusing concave mirrors determine the
minimal collision angle $\alpha$ between laser and electron beam.
They should be kept small for high yield of $\gamma$. This however
gives rise to an additional contribution $LF_{diff}$ to the total
loss factor $V$ of the cavity due to diffractive power loss, to a
diffraction broadening of the focal spot size $w_{CP,\,Gaussian}$
and to deviations from the Gaussian mode. These were studied
numerically using the physical optics code {\sc GLAD}
\cite{GLAD:01}. A Gaussian beam was injected into a cavity formed
by perfect reflecting mirrors and was numerically propagated
according the Fox-Li approach \cite{FL:1961} throughout many
revolutions, until a stationary field distribution was
established. Then the beam size $w_{CP}$ at the focus was
calculated by the second moment of the electric field
distribution. $LF_{diff}$ resulted from the fractional drop of
power during one round-trip.

By reducing the size of the concave mirrors, the beam size
$w_{cc}$ on these mirrors also decreases, minimizing the optical
power that is lost. This requires a growth of the beam waist at
the CP. The relative broadening
$\gamma_{CP}:=w_{CP}/w_{CP,\,Gaussian}$ of the focal spot size
compared to the Gaussian focal beam radius is plotted in
Fig.~\ref{fig:diffrac_broadening}
\begin{figure}[t]
\includegraphics[width=0.9\linewidth,bb=11 0 500 345]{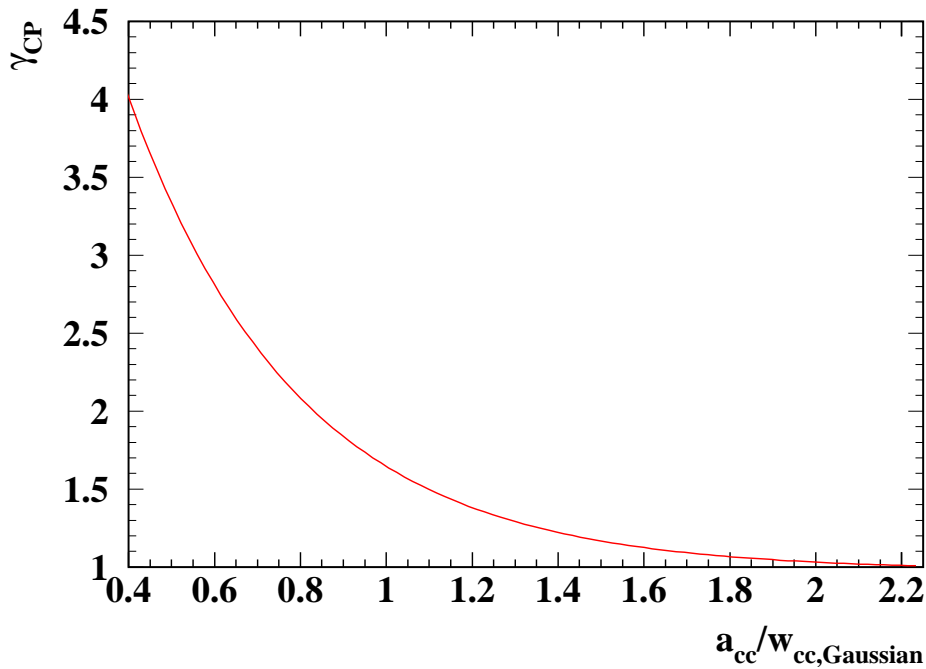}
\caption{Diffraction broadening
$\gamma_{CP}:=w_{CP}/w_{CP,\,Gaussian}$ of the focus at CP caused by
limiting the aperture $2\,a_{cc}$ of the final focusing concave
mirrors. Both concave mirrors are assumed to have the same size.
$w_{cc,\,Gaussian}$ represents the Gaussian beam radius at location
of the concave mirrors that would appear for inifinite large
mirrors. $w_{CP}$ is the corresponding radius for finite aperture.}
\label{fig:diffrac_broadening}
\end{figure}
as a function of a normalized radius $a_{cc}/w_{cc,\,Gaussian}$ of
the concave mirrors. It is expressed in units of the corresponding
Gaussian beam radius at this location. This plot is characteristic
for the layout of an optical cavity following
Fig.~\ref{fig:telescopic_cavity} For reasons of self-resemblance
the diffractive broadening behaves similar when the size of the
focus within the cavity of Fig.~\ref{fig:telescopic_cavity} is
varied by shifting both telescopes either an equal amount towards
(reduction) or away (enlargement) from the CP.

As shown later the diffraction loss $LF_{diff}$ turns out to be
negligible for parameters with acceptable $\gamma_{CP}$.

\section{Laser-electron crossing angle}
\label{sec:lumiopt} In order to calculate the laser-electron
crossing angle $\alpha$ and to specify $w_{CP}$, diffraction
broadening has to be taken into account. In respect to a high
yield of $\gamma$, crossing angle $\alpha$, mirror diameter
$2\,a_{cc}$, waist size $w_{CP}$, laser pulse energy $E_{pulse}$,
as well as laser pulse duration $\tau_{pulse}$ are all
interdependent parameters. Their respective values were determined
by a numerical optimization process using the CAIN Monte Carlo
code \cite{Yokoya:1985ab,CAIN,Chen:1997fz} which assumes charged
particles interact with a Gaussian optical beam. The
center-of-mass energy was set to 500\,GeV. The aperture
$2\,a_{cc}$ of the final focusing concave mirrors $M_4$, $M_5$ at
distance $L_{image}$ from the conversion point CP in
Fig.~(\ref{fig:telescopic_cavity}) sets an upper limit for the
opening angle $\theta_{cc}$ of the laser cone that emerges from
the beam waist:
\begin{equation}
\label{eq:laser_opening_limit2}
\theta_{cc}=\frac{\displaystyle a_{cc}}{\displaystyle L_{image}}
=
\frac{\displaystyle a_{cc}}{\displaystyle w_{cc}}\,
\theta_L
\quad.
\end{equation}
Replacing $L_{image}$ by the far field divergence angle $\theta_L$
originating from a Gaussian beam waist $w_0$ results in the latter
equation. $w_{cc}$ represents the beam radius on each of the
concave mirrors. For the TESLA parameters, an additional offset
angle $\beta$ of 17\,mrad occurs in view of the disruption of the
electron beam during the Compton scattering and the physical size
of the electron beam pipes near the interaction region. The
crossing angle $\alpha$ is thus expressed as
\begin{equation}
\label{eq:crossing_angle}
\alpha = \frac{\displaystyle a_{cc}}{\displaystyle w_{cc}}\,
\theta_L +\beta\ ,
\qquad
\beta=17\,\mbox{mrad}
\quad.
\end{equation}
This relation was encoded into CAIN via the optical Rayleigh
length $z_R=\pi\,w_0^2/\lambda$
 by using the relation $\theta_L=w_0/ z_R$ of Gaussian beams
\cite{Sieg_d:1986}. Instead of the Gaussian beam waist, the
numerical value for the diffraction broadened waist
$w_{CP}$\,=\,$\gamma\,w_{CP,\,Gaussian}$ has been used for $w_0$.
Fig.~\ref{fig:lopt} shows the resulting luminosity
\begin{figure}[t]
\includegraphics[viewport=0 0 490 490,width=0.9\textwidth]{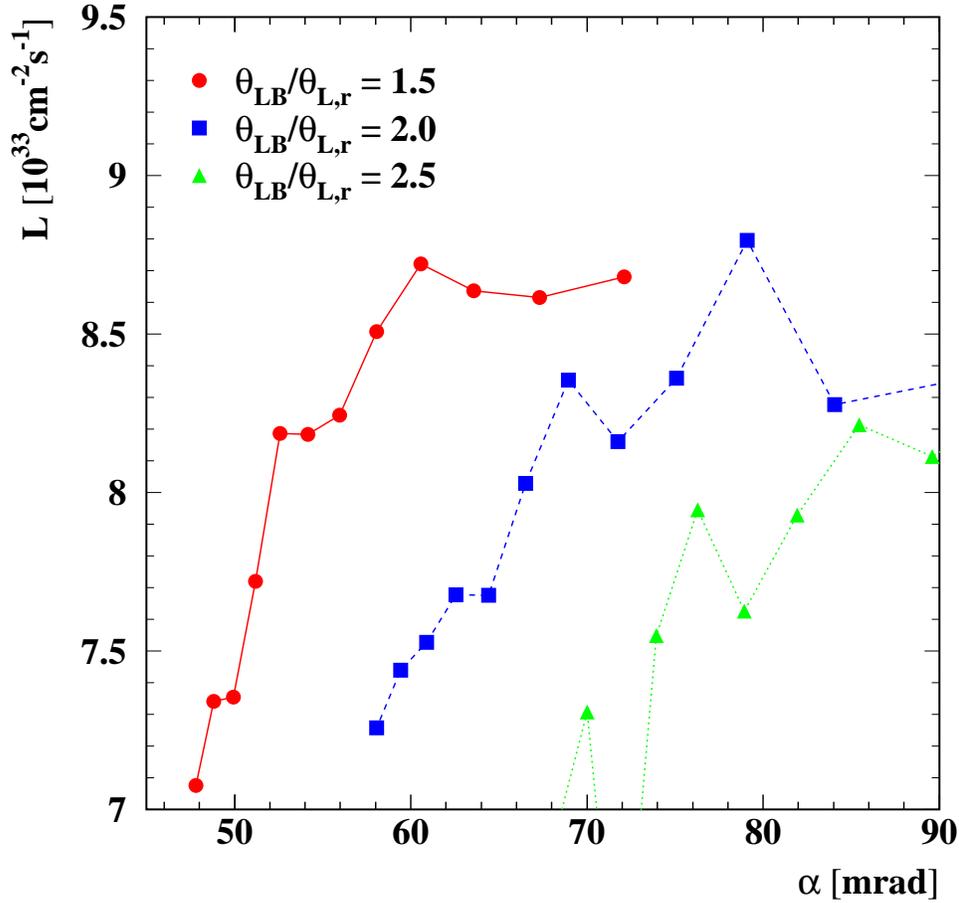}
\caption{The $\gamma\gamma$ luminosity in the high energy part of
the Compton spectrum as calculated using CAIN. It is plotted as a
function of the laser-beam crossing angle $\alpha$ for different
values of the steric opening angle $\Theta_{LB}$ of the concave
mirror. It is given in units of the rms angular divergence
$\Theta_{L,r}$ of the laser light in the focal point
($\Theta_{LB}/\Theta_{L,r}=2\,a_{cc}/w_{cc,\,Gaussian}$). The high
energy part was defined as $z>0.8\,z_{max}$, whereas
$z_{max}=x/(x+1)$, neglecting non-linearity effects.}
\label{fig:lopt}
\end{figure}
as a function of the crossing angle $\alpha$ for different mirror
sizes $a_{cc}/w_{cc}$ and a laser pulse duration of
$\tau_{L}=3.5$\,ps FWHM ($\sigma$\,=\,1.5\,ps). In the examined
range the luminosity rises with decreasing diameter of the
mirrors. A value of $a_{cc}/w_{cc}=0.75$ is therefore selected. An
acceptable crossing angle is then $\alpha\,\approx\,55$\,mrad.
This corresponds to $z_R(w_{CP})$\,$\approx$\,0.63\,mm, a
diffraction broadened beam waist of
$w_{CP}$\,$\approx$\,14.3\,$\mu$m\,($1/e^2$)
($\sigma$\,=\,7.15\,$\mu$m)\, and a Gaussian waist
$w_{CP,\,Gaussian}$\,$\approx$\,6.5\,$\mu$m\,($1/e^2$)
($\sigma$\,=\,3.3\,$\mu$m)\footnote{($1/e^2$) designates the
radius that is defined by a drop of the intensity to
$1/e^2$\,$\approx$\,13.5\,\% of its maximum value at the beam
center.}.

A total luminosity\footnote{Here $z_{max}$ is defined as
$z_{max}=x/(x+1+\xi^2)$ consistent with the definition in
\cite{Telnov:2004}.} of ${\mathcal L}(z>0.8 z_{max}) = 1.1 \cdot
10^{34}\,{\rm cm^{-2}s^{-1}}$ can be achieved for these parameters
with a pulse energy of 9\,J \cite{Bech:03}. A non-linearity
parameter $\xi^2=0.3$ can be maintained in accordance with
reference \cite{Telnov:2004}. In proportion to the laser pulse
energy the required average laser power has also gone up to
9\,J\,$\times$\,2820\,$\times$\,5\,Hz\,$\approx$\,130\,kW. All
resulting parameters for the Compton interaction zone of a
$\gamma\gamma$-collider based on 250\,GeV electron beams are
compiled in Tab.~\ref{tab:Laser-require}.
\begin{table}[t]
\caption{\label{tab:Laser-require} Optical parameters resulting
from an optimization of the $\gamma\gamma$ luminosity}
\begin{tabular}{ll}\hline
laser pulse energy $E_{pulse}$        & $\approx$\,9.0\,J\\
average laser power $<P_{laser}>_t$   & $\approx$\,130\,kW\quad
for one pass
collisions at the\\
& \hspace{11.5ex}TESLA bunch-structure\\[0.5ex]
pulse duration $\tau_{pulse}$         &
  3.53\,ps FWHM ($\sigma$\,=\,1.5\,ps)\\
Rayleigh length $Z_R$                 & $\approx$\,0.63\,mm\\
beam waist $w_{CP}$                   &
  $\approx$\,14.3\,$\mu$m\,($1/e^2$) ($\sigma$\,=\,7.15\,$\mu$m)\\
laser-e$^-$ crossing-angle $\alpha_0$ & $\approx$\,56\,mrad \\
normalized mirror-size $a/w$          & 0.75 \\
laser wavelength $\lambda$            & 1.064\,$\mu$m\\
nonlinearity parameter $\zeta^2$      & 0.30 \\
total luminosity $L_{\gamma\gamma}$   &
  $1.05\cdot 10^{34}$\,cm$^{-2}$s$^{-1}$\\[2ex]
\hline
\end{tabular}
\end{table}
The diffraction broadened beam waist size of
$w_{CP}$\,$\approx$\,14.3\,$\mu$m\,($1/e^2$) corresponds to a
Gaussian beam waist
$w_{CP,\,Gaussian}$\,$\approx$\,6.5\,$\mu$m\,($1/e^2$)
($\sigma$\,=\,3.3\,$\mu$m).

\section{Specification and enhancement capability of the cavity}
In view of the dimension of the particle detector, $t$, $f_c$,
$f_x$ and $\mu$ were selected such that $L_{image}\approx 15$\,m.
The design parameter of the optical cavity according to
Fig.~\ref{fig:telescopic_cavity} are compiled in
Tab.~\ref{tab:geodata}.
\begin{table}[h]
\caption{\label{tab:geodata} Mirror spacings $L$, effective focal
lengths $f$ and folding angles $\gamma_{ijk}$ of the cavity from
fig.~\ref{fig:telescopic_cavity} for 6.5\,$\mu$m Gaussian beam
waist. $L_1$ represents the spacing between $M_4$ and $M_5$ and
$L_2$ the distance between $M_3$, $M_6$ via $M_2$, $M_8$, $M_1$,
$M_7$. $\gamma_{328}$ denotes e.g. the folding angle at mirror $M_2$
enclosed by the distances between ($M_3,\,M_2$) and ($M_2,\,M_8$).
$G$: stability parameter of the cavity.}
\begin{tabular}{l@{\,=\,}l
l@{\,=\,}l l@{\,=\,}l }\hline $L_1$ &3041.85862\,cm \hspace{2ex} &
$f_c$ &809.82\,cm  \hspace{2ex} &
  $t$   &732.05\,cm \\
$L_2$ &5597.05\,cm    \hspace{2ex} & $f_x$ &-1000.0\,cm \hspace{2ex} &
$G^{\ast}+1$
&3.86\,$\cdot$\,$10^{-9}$\\[1ex]
\end{tabular}
\begin{tabular}{llll}
 L(M$_4$, M$_3$) = t           & L(M$_7$, M$_1$) = L(M$_2$, M$_8$)
   \\[0.5ex]
 L(M$_3$, M$_2$) = 530\,cm     & L(M$_7$, M$_6$) = L(M$_3$, M$_2$)
   \\[0.5ex]
 L(M$_2$, M$_8$) = 2083.66\,cm & L(M$_6$, M$_5$) = t \\[0.5ex]
 L(M$_8$, M$_1$) = 369.73\,cm  &
  L(M$_5$, CP) = L(CP, M$_4$) = $L_1/2$\\[1ex]
 $\gamma_{345}=10^{\circ}$  &   $\gamma_{817}=\gamma_{281}$ \\
 $\gamma_{234}=30^{\circ}$ &   $\gamma_{176}=\gamma_{281}$ \\
 $\gamma_{328}=64.63^{\circ}$   & $\gamma_{765}=\gamma_{234}$ \\
 $\gamma_{281}=44.63^{\circ}$ &   $\gamma_{654}=\gamma_{345}$ \\
 beam magnification & $\mu$ = $\sqrt{3}$\\
 diameter of all mirrors & 120\,cm\\[0.5ex]\hline
\end{tabular}
\end{table}
The mode within the optical cavity produces its largest spot-size
at the concave mirrors next to the beam waist. In a hypothetical
cavity with mirrors of unlimited size a Gaussian beam of
$w_{cc,\,Gaussian}$\,=\,79.1\,cm radius ($1/e^2$) would develop.
Assuming a clipping aperture of 75\,\% of the Gaussian beam
radius, i.e.
$a_{cc}$\,=\,0.75\,$\cdot$\,$w_{cc,\,Gaussian}$\,=\,59.3\,cm
$\approx$\,60\,cm, the $1/e^2$ spot size on the concave mirrors
reduces according to the numerical modelling to
$\approx$\,42.7\,cm. This leaves sufficient room for the wings of
the electric field distribution as will be seen in the discussion
of diffraction loss. The concave mirror should therefore have a
diameter of about 120\,cm. This yields a f-number
$f_{\#}$\,=\,12.7 for both focussing telescopes.

In this way the concave mirror is designed to represent the
dominant aperture at which diffraction will occur. The telescope
reduces the lateral extension of the mode by a factor of
$\sqrt{3}$. However, a reduction of all other mirrors by this
factor $\sqrt{3}$ is not advisable without introducing further
significant apertures. These would influence the beam size even
further and create losses which in the end would reduce the
efficiency of the Compton conversion. Therefore, all mirrors
should have the same diameter of 120\,cm.

A plot of the numerically obtained stationary diffraction loss
factor $LF_{diff}$ as a function of the diffraction broadened beam
waist $w_{CP}$ is shown in Fig.~\ref{loss-finite-aperture}.
\begin{figure}[t]
\includegraphics[width=0.9\textwidth]{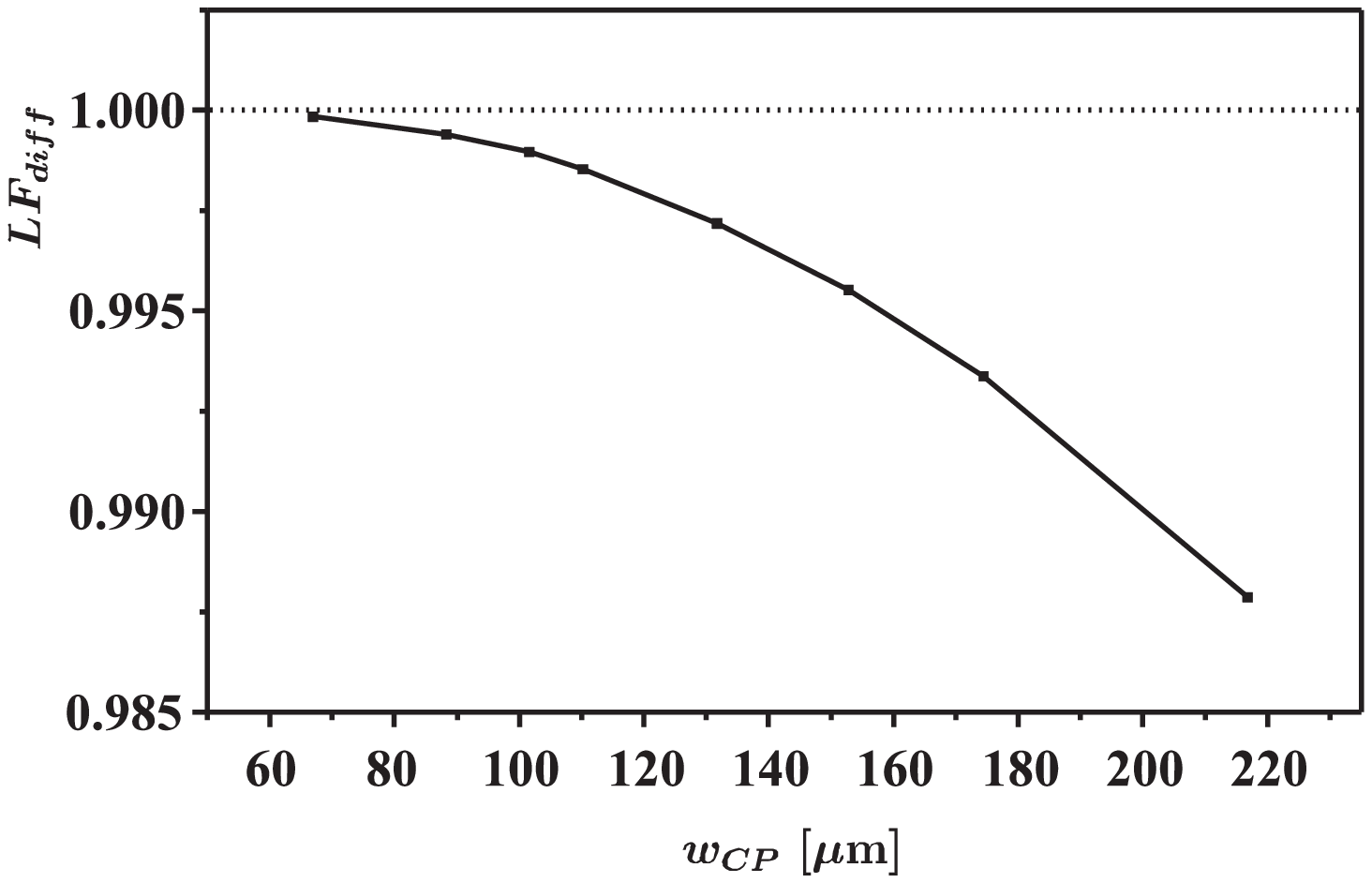}
\caption{Numerically obtained stationary diffraction loss factor
$LF_{diff}$ for a set of cavities with finite apertures of all
mirrors for a sequence of focal spot sizes $w_{CP}$ at the Compton
conversion point. The scaling of the apertures has been chosen
such that the ratio $a_{cc}/w_{cc,\,Gaussian}=0.75$ is kept as for
the optimized cavity. Each time, the aperture of the remaining
mirrors has then been set equal to that of the concave mirrors.}
\label{loss-finite-aperture}
\end{figure}
It results from slight variations of the distance $L_1$ between
both telescopes. At the same time $L_2$ has been adjusted in the
opposite direction, in order to keep the total path length
unchanged. The diameter of all mirrors was thereby adapted to end
up with the same ratio $\,a_{cc}/w_{cc,\,Gaussian}=0.75$ as for
the optimized cavity. Diffraction loss declines according to
Fig.~\ref{loss-finite-aperture} towards smaller foci within the
cavity. $LF_{diff}$\,=\,1 denotes no power loss due to
diffraction.

A diffraction loss factor of roughly $LF_{diff}\ge 0.9998$ was
obtained from an extrapolation down towards $14.3\,\mu$m
diffraction broadened beam waist. Taking the reflectivity of
practical highly reflecting mirrors into account reduces the total
loss factor to $V=\,LF_{diff}\,{R_{HR}}^7$. $R_{HR}$ denotes the
reflectivity of all remaining mirrors with exception of the
coupling mirror. A reflectivity between $R_{HR}$\,=\,99.99\,\% and
99.95\,\% would permit a steady-state impedance matched power
enhancement between 1100 and 270, for otherwise perfect
conditions. The enhancement becomes the more sensitive against any
impedance mismatch, the larger the amount of $A$ is. In practice,
one probably will have a set of spare mirrors with slightly
varying reflectivity and check which one will give the best
enhancement.

Before onset of each electron bunch train $n_{\kappa,\,prepulse}$
additional laser pulses are required for the accumulation of a
sufficient large optical pulse energy within the cavity, that will
ensure an enhancement of at least the fraction $\kappa$
(${\kappa}<1$) of the stationary amount $A_{max}$. The number
$n_{\kappa,\,prepulse}$ is derived from Eq.~(\ref{resonant_Pcirc})
as follows:
\begin{equation}
\label{eq:prepulse_number} n_{\kappa,\,prepulse}=
\frac{\displaystyle \ln\left(1-\sqrt{\kappa}\right)}{\displaystyle
\ln(\sqrt{R_c\,V})} \quad.
\end{equation}
As an example, the build-up of power obtained from the numerical
propagation of an optical pulse under Gaussian seed-conditions
through many revolutions is represented by
Fig.~\ref{fig:numerical_enhancement} for $R_c$\,=\,99\,\%,
V\,=\,0.9998, and $R_{HR}$\,=\,100\,\%. At perfect resonance 1034
pre-pulses are then expected for obtaining an enhancement of at
least 99\,\% of $A_{max}$\,=\,383, i.e. $A$\,=\,378.8. The
stationary enhancement of 353 in
Fig.~\ref{fig:numerical_enhancement} following in accordance with
Eq.~(\ref{eq:prepulse_number}) after $\approx$\,1000 circulations of
the pulse reflects a 92\,\% coupling of the injected Gaussian mode
into the cavity mode. The number of laser pre-pulses for reaching an
approximate steady-state declines as the impedance matching
condition is violated.
\begin{figure}[t]
\includegraphics[width=0.9\textwidth]{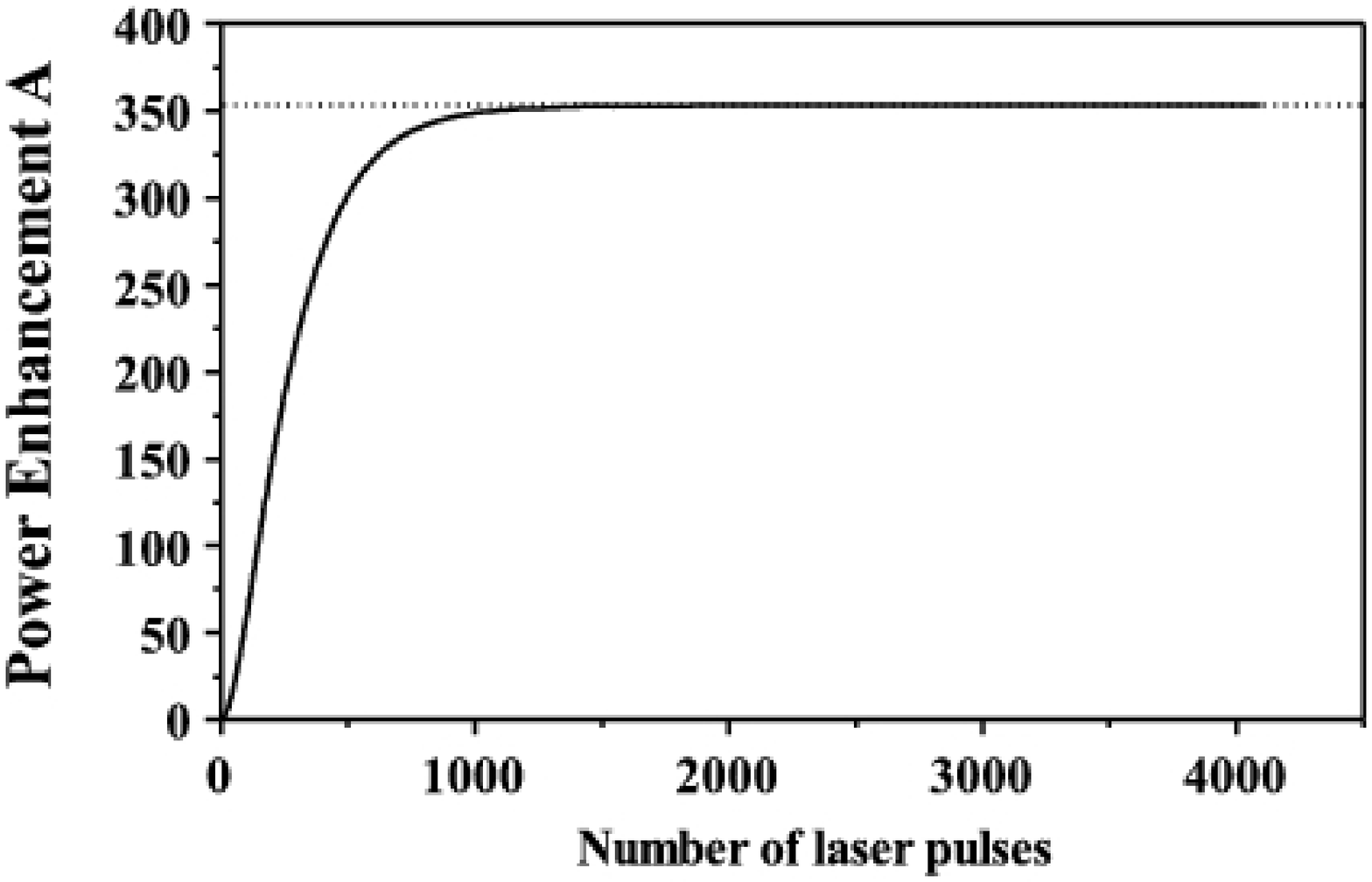}
\caption{Representative numerical example for power build up,
assuming a total loss factor $V$\,=\,0.9998 and a reflectivity of
$R_c$\,=\,99.0\,\% for the coupling mirror. This demands the
accumulation of at least 1034 pre-pulses for reaching an
enhancement within $\kappa$\,=\,99\,\% of the steady-state. For
details see text.}
\label{fig:numerical_enhancement}
\end{figure}

When $A$ represents the power enhancement factor of the optical
cavity, the required average power of the laser is reduced to
$A/(1+n_{prepulse}/n_{train})$, where $n_{train}$ is the number of
electron bunches in one train.

\section{Damage threshold}
The TELSA bunch structure consists of bunch trains of 2820 bunches
with about 300\,ns bunch spacing and a train repetition rate of
5\,Hz. This pulse structure represents an intermediate regime
between two identified damage mechanisms \cite{Koe:1999}. For a
pulse duration above 100\,ps the damage occurs by melting due to
heat deposition, whereas for less than 20\,ps the damage site is
limited to the region where the intensity is sufficient for a
laser generated plasma. This occurs before a significant transfer
of energy from the electrons to the lattice has taken place.

For $\approx$\,3.5\,ps (FWHM) pulse duration the threshold for 600
shots at 10\,Hz repetition rate and 1053\,nm wavelength of
uncoated very uniform fused silica samples is of the order of
2\,J/cm$^2$ \cite{SFH:1996}, whereas just $\approx$\,5\,mJ/cm$^2$
per pulse results from the parameter for the cavity.

For a rough estimation of the worst case risk for material damage
the cumulative effect of all pulses within the train is assumed to
be represented by a single pulse with total energy and duration
equivalent to that of all optical bunches within a train. The
parameters of Tab.~\ref{tab:Laser-require},\,\ref{tab:geodata}
result then in a laser energy fluence of around 4\,J/cm$^2$ within
10\,ns at the mirrors of the cavity for a hypothetical Gaussian
beam and 13\,J/cm$^2$ for the cavity mode with truncated mirrors.
At the final focussing concave mirrors this fluence is lower by a
factor of 3 due to the beam expansion. According \cite{Koe:1999}
the laser damage threshold for 10\,ns pulse duration for various
materials including fused silica and coatings is higher and lies
around 50\,J/cm$^2$ for the substrat, and above approximately
44\,J/cm$^2$ to 120\,J/cm$^2$ for the mirrors depending on the
composition of the multilayer coatings. Following the known
scaling laws summarized e.g. in \cite{Koe:1999}, for a wavelength
around 1\,$\mu$m a further factor of
$(1\,\mbox{ms}/10\,\mbox{ns})^{0.4} = 100$ for the increase of the
threshold for the contribution due to thermal induced damage could
probably be anticipated for distribution of the optical bunches
within a train of $\approx$\,1\,ms duration as in the TESLA time
structure.

This means that the expected fluence is below the damage
threshold. However, no data for trains of ps-pulses separated on
nano- to microsecond time scales which accumulate to the stated
fluences are known to us. For a final judgment an experimental
study with a representative of the ILC bunch structure would be
required.

\section{Effects of cavity misalignments}
>From Eq.~(\ref{Ehf}) the maximum acceptable length detuning
$\delta_{\kappa}$ for the circumference of the optical cavity for
a tolerated fractional decline $\kappa$ in the power enhancement
factor $A_{max}$ from the resonance condition is derived to be
given by
\begin{equation}
\label{eq:longitudinal-criterion}
\delta_{\kappa} \approx
\frac{\lambda}{2\,\mathcal F} \sqrt{\frac{1}{\kappa}-1}
\qquad(\mathcal F \gg 1)
\qquad.
\end{equation}
Maintaining the power enhancement factor e.g. above
$\kappa=90$\,\% of its optimum value $A_{max}$ demands a match of
the circumference of the cavity of better 0.57\,nm for an assumed
value $A_{max}$ of 100. Technical solutions for such a precise
length stabilization are well-known \cite{DH:1983}. Even more
stringent restrictions are common in interferometrical detection
of gravitational waves.

Any misalignment generally results in displacement and broadening
of the intra-cavity beam waist. Mechanical vibrations due to
instrumentation in the environment of the particle detector,
seismic ground motion as well as slight deviations during assembly
might slightly modify position and tilt of the mirrors. According
to our calculations, the displacement of the beam waist remains
smaller then the Rayleigh length, i.e. the depth of the focus.
This shift of the beam waist is hence negligible.

Broadening of the waist modifies the cavity-mode and reduces the
mode coupling between laser and cavity. The resulting relative
decline of the power enhancement factor $A_{max}$ was found to be
largely independent of the finesse of the cavity. As an example
two identical cavities with enhancement factors of 353 and 76,
respectively, were considered. Fig.~\ref{fig:misalignment_a} shows
the effect of shifting one telescope
\begin{figure}[t]
\includegraphics[width=0.9\linewidth]{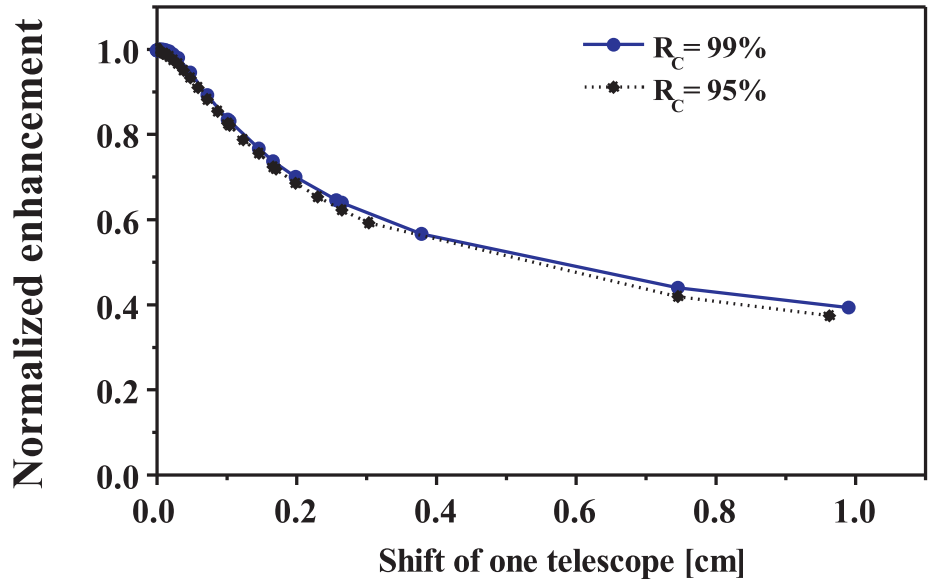}
\caption{ Calculated effect of an axial misalignment of one
telescope on the power enhancement for two different values of the
coupling reflectivity $R_C$ and a mirror size adapted according
$a_{cc}/w_{cc,\,Gaussian}=0.75$. A cavity loss factor
$V$\,=\,0.99938 was arbitrarily assumed. Then the coupling
reflectivities $R_c$\,=\,95\,\%\ and 99\,\% result in
$A_{max}$\,=\,76 and 353, respectively, for the undisturbed
cavity. At zero shift these values have been scaled to $1$ for
comparison of their fractional decline.}
\label{fig:misalignment_a}
\end{figure}
over a range of up to 1\,cm from its designated location away from
the waist. Its internal alignment was assumed as being preserved.
Regarding this vast variation range compared to the Rayleigh
length in Tab.~\ref{tab:geodata}, the decline for sub-mm shifts is
rather modest. It remains below 20\,\%. At a shift of 1\,cm
reduces the enhancement to 40\,\% of $A_{max}$. A decline of
similar amount occurs e.g. for a deviation of the focal length
$f_x$ of the convex mirror of 12\,mm.

An increase of the focal length $f_x$ or shift of the telescope
away from CP (the location of the beam waist) is accompanied by an
increase of the radius of the beam waist. The latter is depicted
by Fig.~\ref{fig:misalignment_c}. Under the influence of clipping
at the concave mirrors the growth of the diffraction broadened
beam size is considerably lower. This allows to shift the
operating point of the cavity further away from the stability
limit. A zero waist size denotes an instable optical cavity.
\begin{figure}[t]
\includegraphics[width=0.9\linewidth]{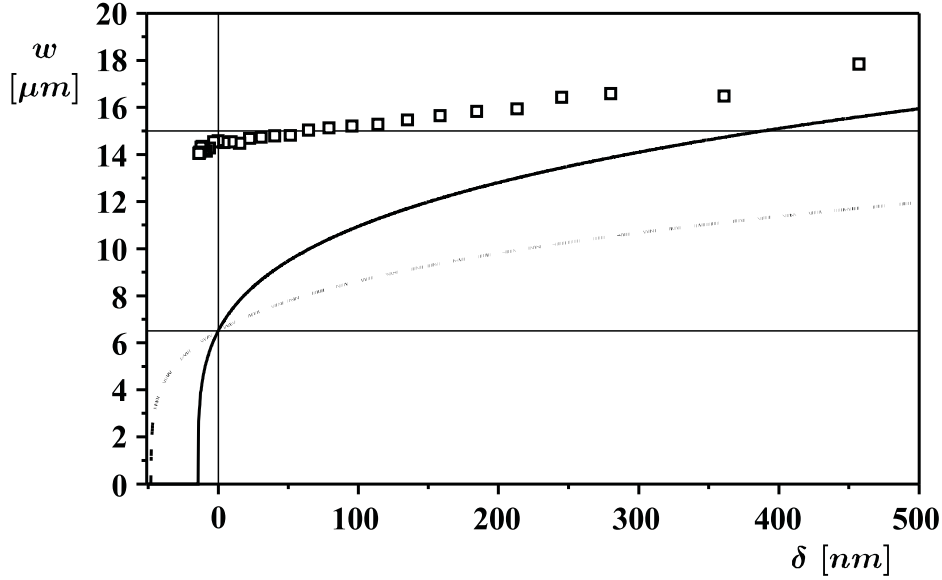}
\caption{ Sensitivity of the waist within a cavity for nominal
6.5\,$\mu$m\,($1/e^2$) Gaussian waist against axial displacement
$\delta$ of either one concave (dashed) or convex mirror (solid
line), as well as the corresponding waist for a mirror size scaled
according $a_{cc}/w_{cc,\,Gaussian}=0.75$ (squares). For
$\delta$\,=\,0 the beam radius increases to the required $\approx
15\,\mu$m\,($1/e^2$).}
\label{fig:misalignment_c}
\end{figure}

\section{Discussion}
The $\gamma\gamma$-collider will be operated with trains of
optical ps-pulses, whereas the optical interferometric detection
of gravitational waves relies on continous-wave (cw) laser
radiation. An automatic alignment system has already been
developed for gravitational wave detection \cite{GHF:02,HRS:01}.
Therefore, use of an additional uninterrupted train of weak
ps-pulses for generation of error signals resulting from
misaligned components and constant control of the cavity seems
promising, as the cavity for the $\gamma\gamma$-collider can
benefit from that knowledge.

The tiny absorption of the laser beam within the substrate and
mirror coating is expected to induce a nonuniform temperature
increase within the optic due to the amount of circulating optical
power. This causes a nonuniform distortion of the optical path
length by thermal expansion of the optic's surface, and a
variation of the material's refractive index with temperature
(thermal lensing). In our numerical model the focal length has to
stay within a few tenth mm of its exact value for missing the
desired beam waist by not more than 5\,\%. A correction of the
curvature radius of imaging mirrors has already been demonstrated
by exerting an axially symmetrical mechanical strain within the
reflecting surface through radiative heating \cite{Law:02}. The
remaining non-axial-symmetric wavefront distortion generated by
inhomogeneities in the substrate can be compensated by locally
heating the mirror in addition with another laser via further
computer controlled scanning mirrors. A proof-of-principle
experiment has been performed in \cite{LMAD:01}.

Moreover, the high misalignment sensitivity of the proposed cavity
could be overcome by the introduction of adaptive optics for
ensuring and controlling the development of a diffraction limited
optical mode. This requires an additional control loop acting on
the surface shape of at least one deformable mirror. The flat
folding mirrors M$_1$ and M$_2$ adjacent to the telescopes in
Fig.~\ref{fig:telescopic_cavity} appear to be especially well
suited for this task.

The proposed cavity approximates a half-degenerate cavity which is
characterized by a round-trip matrix of -1 times the unity matrix.
In a half-degenerate cavity, the intensity distribution at any
position along the propagation axis is relay imaged at completion
of two circulations. Such a cavity can already be described in
good approximation by ray-tracing (geometric-optical imaging).
Diffraction has only be taken into account for calculation of the
focus.

\section{Conclusion}
The actual enhancement factor of the cavity results from the
cumulative effect of many small contributions affecting the total
loss factor. Residual abberations tend to enlarge the focused beam
at the Compton conversion point. Both are difficult to predict in
advance. However, the influence of diffraction loss is for the
proposed cavity almost negligible.

According to an estimation based on the known properties of laser
induced damage, the suggested size of the mirrors provides still
some reserve before the energy fluence of the circulating optical
pulse reaches the damage threshold. For a final judgement of the
upper limit, an experimental study with a representative of the
special ILC bunch structure is required.

The use of adaptive optics appears to be essential for operation
of such a cavity.

\section{Acknowledgements}
We would like to thank J.~Gronberg, I.N.~Ross, A.~Stahl, and
V.~Telnov for many useful discussions. We are also grateful to
F.~Bechtel for his help during numerical calculations of the
luminosity optimization as well as to N.~Meyners of DESY's MEA
department for preparing the steric schemes used in
Fig.~\ref{fig:embedded-cavities}.

\end{document}